# Quantifying Efficiency of Remote Excitation for Surface Enhanced Raman Spectroscopy in Molecular Junctions

Shusen Liao[1,2,†], Yunxuan Zhu[2,†], Qian Ye[2], Stephen Sanders[3], Jiawei Yang[2], Alessandro Alabastri[3], Douglas Natelson[2,3,4*]

[1] Applied Physics Graduate Program, Smalley-Curl Institute, Rice University, Houston, TX 77005 USA

[2] Department of Physics and Astronomy, Rice University, Houston, TX 77005 USA

[3] Department of Electrical and Computer Engineering, Rice University, Houston, TX 77005 USA

[4] Department of Materials Science and NanoEngineering, Rice University, Houston, TX 77005 USA

[†] These authors contributed equally to this work.

## Abstract

Surface-enhanced Raman spectroscopy (SERS) is enabled by local surface plasmon resonances (LSPRs) in metallic nanogaps. When SERS is excited by direct illumination of the nanogap, the background heating of lattice and electrons can prevent further manipulation of the molecules. To overcome this issue, we report SERS in electromigrated gold molecular junctions excited remotely: surface plasmon polaritons (SPPs) are excited at nearby gratings, propagate to the junction, and couple to the local nanogap plasmon modes. Like direct excitation, remote excitation of the nanogap can generate both SERS emission and an open-circuit photovoltage (OCPV). We compare SERS intensity and OCPV in both direct and remote illumination configurations. SERS spectra obtained by remote excitation are much more stable than those obtained through direct excitation when photon count rates are comparable. By statistical analysis of 33 devices, coupling efficiency of remote excitation is calculated to be around 10%, consistent with the simulated energy flow.

**TOC graphic**:

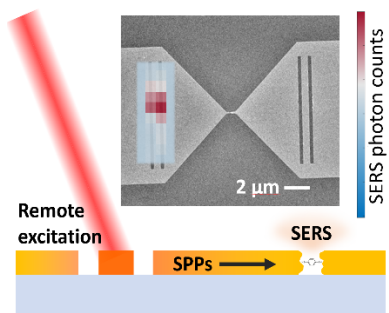

[*] <natelson@rice.edu>

Raman scattering is the inelastic scattering of light by matter and was first discovered by the Indian physicist C. V. Raman and his student K. S. Krishnan.[1] In Raman scattering, the incoming light can interact with molecular motions (vibrations, rotations) that alter the polarizability tensor of the molecule. The incident photon can either lose or gain energy in the interaction, leading to Stokes or anti-Stokes Raman scattering, respectively. The photon energy shift in Raman scattering can be used to determine the vibrational modes of the molecules, so it provides a fingerprint of molecular systems, and Raman spectroscopy is a powerful tool for molecule identification in chemistry, biology and other related fields.[2–7] However, the intensity of Raman spectroscopy can be low, especially when the number of the molecules is small, due to the small scattering cross section. One solution to this problem is surface-enhanced Raman spectroscopy (SERS).[8–11]

SERS typically occurs when the molecules are adsorbed on rough metal surfaces and nanostructures. Electromagnetic theory can explain one major, widely accepted mechanism of SERS. Incident light with electric field amplitude $E_0$ can excite LSPRs in the metallic nanostructures.[12] As a result, hot spots where the local electric field $E$ is dramatically enhanced are generated. The SERS signal from the hot spots can be enhanced roughly by a factor of $g^4$, where $g = \frac{E}{E_0}$ is the electric field enhancement.[13,14] Additionally, there can be a "chemical enhancement" contribution to SERS that is related to charge transfer and resonance Raman scattering.[15–17] In experiments, the SERS enhancement can exceed $10^{10}$, which means it reaches the single molecule detection region.[18–20] Because of its high sensitivity, SERS is widely used for many chemical and biological applications,[11,21,22] such as *in situ* monitoring of chemical reactions[23].

SERS has been reported in many plasmonically active metal nanostructure environments, for example, nanoparticles,[24] structured nanofilms,[25] and nanoscale tips in scanned probe microscopes.[26] Planar Au nanowires can be electromigrated to form a nanogap[27], and molecules can be deposited inside and around the gap to obtain a molecular junction. These junctions can be designed to have good plasmonic properties in the near infrared region, can be fabricated on Si substrates on large scales, and support SERS.[28,29]

Localized gap plasmon modes can be excited by the incident light and the electric field inside the gap can be greatly enhanced over the field of the incident radiation,[30] causing SERS signals from these hot spots. In addition to SERS, other plasmon-related behaviors are also observed in the nanowire junction system. Plasmons can decay non-radiatively to electron and hole pairs through Landau damping.[31–33] When the electrons and holes are driven out of the thermal equilibrium distribution, they are called hot carriers.[31] Due to the random shape of the gap after the electromigration, hot electrons can be generated asymmetrically across the nanogap, leading to an optically driven hot carrier tunneling current. For the net current to be zero in an open-circuit configuration, an open-circuit photovoltage (OCPV) must be generated to balance the hot electron tunneling current.[34]

The easiest way to optically excite the nanogap system for SERS as well as OCPV is to shine a laser directly at the gap. In the direct excitation configuration, however, local heating of the metal is substantial, with the lattice temperature rise of the nanowire reaching ~100 K with 220 kW/cm$^2$ incident laser intensity when the substrate temperature is low.[35] This greatly hampers potential further manipulation of the molecules, such as electrostatic gating and inelastic electron tunneling spectroscopy.[36,37]

Remote excitation for SERS has been demonstrated in recent years in nanorods and tip-based systems.[38,39] A laser is focused on a location which is several micrometers away from the SERS hotspots and molecules. Surface plasmon polaritons (SPPs) are generated and propagate to the molecule position to excite SERS signals. To date, the efficiency of this remote SERS approach has not been examined in detail. In the planar nanogap system, remote excitation gives a temperature rise at the nanogap expected to be a factor of 60 smaller than direct excitation.[40] OCPV by remote excitation in the electromigrated junctions is also observed.[41] Here, we report SERS by remote excitation in nanogap-based molecular junctions and quantify its efficiency relative to direct excitation of the same nanogaps. With laser polarization perpendicular to the grating slits, SPPs are generated at the gratings which are about 5 μm away from the junction. SERS signals by both remote and direct excitation are measured for an ensemble of devices, and the coupling efficiency of remote Raman excitation into the devices is found to be about 10% relative to direct excitation, which agrees well with simulation results. By comparing the time variation of the spectra, remote SERS spectra are found to be much more stable than direct excitation SERS, presumably thanks to less thermal heating and more stable junction configurations. Our results provide insight of the plasmonic properties and the energy flow in the molecular junction system.

All the devices are fabricated on Si wafers with 2 μm thick thermal oxide $SiO_2$ layer, as shown in Figure 1A. The width of the Au nanowire is designed to be 120 nm and the length is 800 nm, between larger contact pads. The nanowire width establishes a strong transverse plasmon resonance near 785 nm, the wavelength of the Raman excitation. To enable coupling of remote excitation to SPPs, the larger pads include gratings consisting of two slits. The grating center is 5.9 μm away from the nanowire center. The slits are 8 μm long and 250 nm wide, and the distance between the slits is 500 nm. The thickness of Au is 30 nm. The above parameters for the geometry of the device are selected based on the previous work.[40] The devices are wire bonded after benzene-1,4-dithiol (BDT) molecule deposition via self-assembly from solution. The nanowire is electromigrated to form the nanogap junction as described previously.[27] See Supporting Information Sect. S1 for more details on device fabrication.

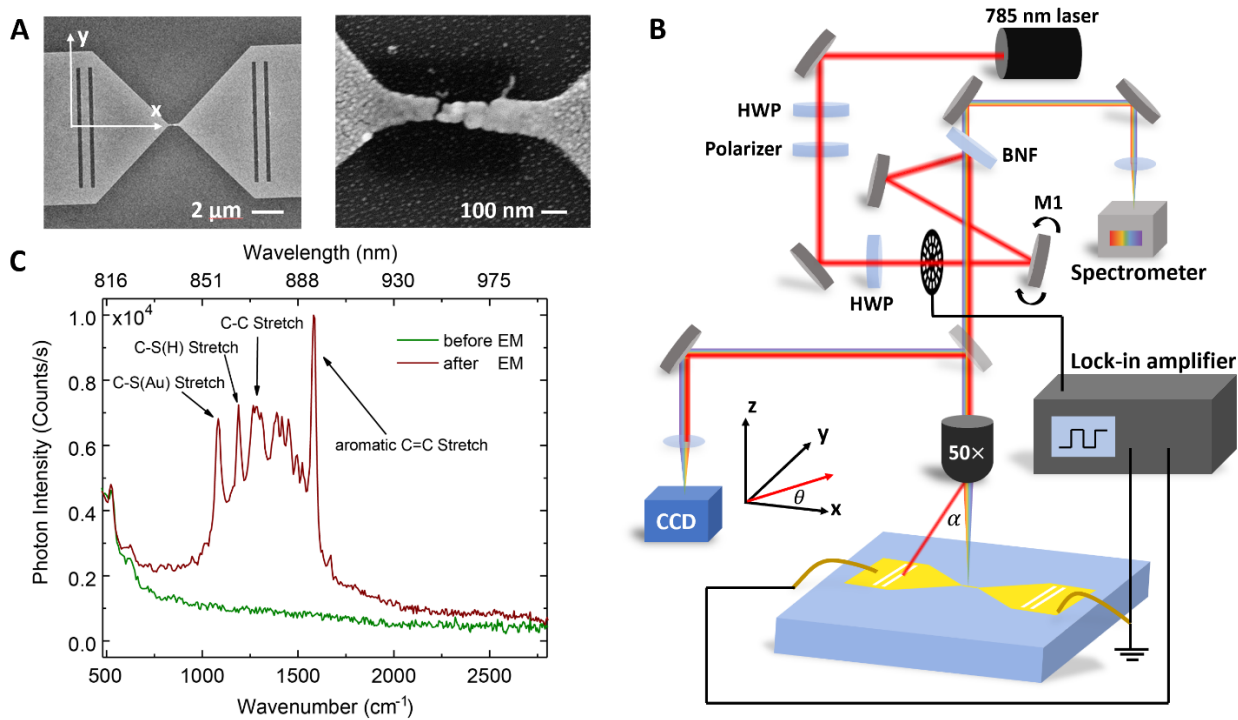

**Figure 1.** (A) Scanning electron microscopy (SEM) images of the device after migration, with gratings apparent in the larger pads. Directions are defined as shown, with illumination polarization along *x* corresponding to 0 degrees and along *y* corresponding to 90 degrees. Left panel: low magnification image of the gratings and nanowire. Right panel: high magnification image of the nanogap. (B) Scheme of the combined SERS and OCPV measurement setup. HWP: half-wave plate. BNF: Bragg Notch filter. The excitation polarization angle is $\theta$. (C) SERS spectra collected from the nanogap when the gratings are subjected to remote $\theta = 0$ excitation before (green) and after (dark red) the electromigration. Four Raman modes of 1-4 BDT are assigned in the post-electromigration spectrum.

The measurement setup for the OCPV and SERS by remote excitation is shown in Figure 1B. Emission from a 785 nm continuous wave (CW) laser is modulated by the optical chopper and focused on the sample at the desired grating with a small angle $\alpha$ shift from the normal incidence controlled by the mirror M1.[42] The outcoming SERS signal from the nanogap also enters the same objective at a different angle and is measured by the spectrometer. Any SERS signal generated by direct excitation at the location of the grating is blocked by the entrance slit of the spectrometer due to the angle difference. For direct excitation, M1 is tilted to place the normal incidence laser at the junction. The OCPV is measured by the lock-in amplifier synced to the chopper. More details about measurements are shown in the Supporting Information Sect. S2. The SERS signals by remote excitation before and after electromigration are measured, as shown in Figure 1C. After electromigration, we can see several Raman modes of the BDT molecule between 1000 to 1600 cm$^{-1}$. Some Raman modes are overlapping, but we can still resolve four Raman modes assigned according to previous reports.[43–45] Before electromigration, there are no detectable Raman modes in this range, consistent with the signal originating from the junction once the nanogap hotspot is created by electromigration.

We obtain maps of OCPV and SERS signal by scanning the excitation laser spot with user-defined pixel size over the junction for direct excitation and over the grating area for remote excitation. The OCPV $V_{OC}$ and SERS response are measured at each pixel. The SERS signal is always collected from the junction by the objective no matter the position of the excitation spot. The results are shown in Figure 2. To quantify SERS response for these maps, all the Raman photon counts are summed in the 1000 to 1600 cm$^{-1}$ range. As the OCPV and SERS are known to change linearly with the input optical power when the input optical power is low (lower than 100 μW/ μm$^2$),[34,46–48] both the OCPV and Raman count rate are normalized by the incident laser power.

We note that the OCPV is an additional benchmark for judging the degree to which incident optical energy is concentrated into the nanogap LSPRs. Strong excitation of nanogap LSPRs is essential to generate a substantial OCPV, just as such LSPR excitation is needed for SERS response. There is not a simple, direct correlation between OCPV and SERS response. Indeed, the magnitude and even the sign of the OCPV under remote excitation can vary substantially[41]. Since both OCPV and SERS are driven by the nanogap LSPRs, statistically we expect to see the devices with larger hot electron current to tend to have stronger SERS signals. Considering each single device, however, the magnitude of the OCPV depends on the asymmetry of the nanogap and the particular LSPR modes. For example, we should have very small

OCPV for devices with highly symmetric gap geometry because the tunneling current from the two sides of the gap approximately cancel, even if the plasmon excitation is strong. However, we may still see strong SERS signal in such a device due to the large electric field enhancement by plasmon resonance. More discussion about OCPV – SERS correlation and the relationship of OCPV to heating is presented in the Supporting Information in Sect. S7 and Figure S11.

When the incident laser is polarized in the 90 degree orientation ($y$-direction, transverse to the nanowire and parallel to the grating slits, $\theta = 90$), we can see very strong OCPV and SERS by direct excitation of the nanogap but there is almost no signal when the laser spot is positioned for remote excitation of the grating. Conversely, when the laser is polarized in the 0 degree orientation ($x$-direction, parallel to the nanowire and transverse to the grating slits, $\theta = 0$), we can see a much weaker OCPV and SERS by direct excitation of the nanogap, but the signal is strong when the laser is positioned on the grating for remote excitation.

These observations are consistent with expectations: the 90 degree polarized light, by direct excitation, can couple to the transverse dipolar plasmon mode of the nanowire, which is hybridized with high-order LSPR modes localized at the nanogap due to the asymmetric geometry of the junction.[49] As a result, a stronger plasmon resonance is generated for this incident polarization. This has two consequences. The efficient excitation of the nanogap LSPRs by direct illumination of the nanogap generates a large number of hot electrons via plasmon decay, leading to a larger OCPV[34] (Figure 2A, left panel). The stronger LSPR response also leads to a highly localized and dramatically enhanced electric field, which contributes to the larger SERS signal[50] (Figure 2B, left panel).

Conversely, for remote excitation, the 90 degree polarized light cannot couple efficiently to the grating, and very little energy is transferred to the SPPs. Thus, both the remote OCPV (Figure 2A, right panel) and integrated SERS signals (Figure 2B, right panel) are small. We note that the dominance of the nanowire transverse plasmon mode for direct illumination, as described above, is quite sensitive to nanowire width and thickness. Over the whole ensemble of devices, by direct excitation, about 70% show that both SERS and OCPV are larger with 90 degree polarized light, in Figure 2. The remaining 30% of devices under direct illumination of the nanogap have maximum OCPV and SERS responses closer to 0 degree incident polarization. This variation can be explained by the several nanometers Au thickness variation and roughness as well as device-to-device variation of the gap geometry. More discussion is shown in the Supporting Information Sect. S5.

When the laser is polarized in the 0 degree orientation, however, although the coupling to the nanowire transverse plasmon mode is weak, the high order LSPRs localized at the gap can be still be weakly excited by direct illumination. This weaker plasmon response still results in some hot carrier tunneling and localized electric field enhancement, so we can still see the OCPV (Figure 2C, left panel) and SERS (Figure 2D, left panel) signals by direct excitation. For remote excitation, 0 degree polarized light can couple efficiently to the grating and generate SPPs which then propagate to the junction, couple to the nanogap LSPRs and create hot electrons and enhanced fields, so we have both relatively large remote OCPV[41] (Figure 2C, right panel) and SERS (Figure 2D, right panel) signals. The propagating SPPs are the summation of SPPs at Au-vacuum and Au-$SiO_2$ interface. Additional details of theoretical and simulated results are shown in the Supporting Information Sect. S3.

It is important to consider the factors that affect the efficiency of remote excitation SERS relative to direct excitation SERS. Qualitatively, during the remote excitation process, there are some energy losses when the light couples to the grating, in the excitation of the propagating SPPs, in their propagation to the junction region, and in their coupling to the nanogap LSPRs. Due to these inefficiencies, the OCPV and SERS by remote excitation are much smaller than those by direct excitation for a fixed laser power. The OCPV and SERS signals by direct excitation are highly localized at the pixel containing the gap, no matter the polarization of the laser. The remote signals are largest when the laser spot is positioned at the middle area of the grating because the coupling efficiency between the grating and the incoming laser, and between the grating and the propagating SPP modes, is higher in that configuration. The SERS spectra measured at different pixels at the grating by remote excitation are shown in Figure 2E. The Raman modes positions and relative strength are the same; the only difference is the magnitude of Raman intensity, as moving the laser away from the middle of the grating degrades the efficiency of coupling energy to the nanogap.

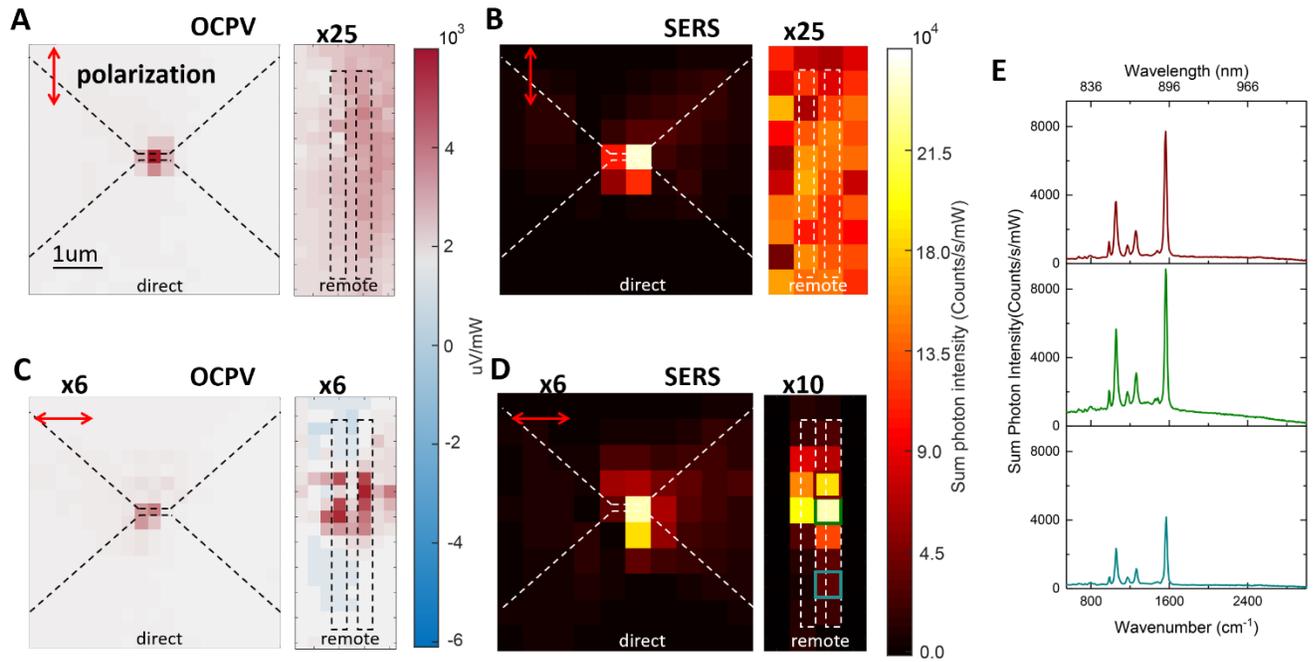

**Figure 2**. Map scans of OCPV and SERS by direct and remote excitation with different input laser polarization. For each map, the excitation laser is located at each pixel position and the relevant response is measured. The square maps correspond to direct excitation (nanogap directly illuminated during the scan), while the rectangular maps correspond to remote excitation (gratings illuminated during the scan). SERS signal is always collected from the junction by the objective for both direct and remote excitation. (A) OCPV with 90 degree polarized light. (B) SERS with 90 degree polarized light. (C) OCPV with 0 degree polarized light. (D) SERS with 0 degree polarized light. (E) SERS spectra with the excitation positioned at the pixels indicated by different colors in (D).

By measuring an ensemble of 33 devices, we perform a quantitative statistical analysis of our system. SERS by both direct and remote excitation are measured for each device. The coupling efficiency of the remote SERS is defined to be the ratio between photon count rates by remote and direct excitation at fixed incident laser power. As shown in Figure 3A, the coupling efficiency averaged across all the devices for each Raman mode is found to be around 10%. For most individual devices, the coupling efficiency averaged across all the Raman modes is also about 10% (Supporting Information Sect. S7). This suggests the 10% coupling efficiency is the intrinsic property of our device geometry design.

Electrodynamic simulations reveal one of the dominant mechanisms limiting the efficiency of remote excitation SERS. The simulated energy propagation in the +$x$-direction from the grating to the junction is shown in Figure 3B, calculated by 3D finite element simulation of the full device geometry and plotting at each $x$ location the $x$-directed Poynting flux integrated over the transverse cross-section. The simulated power of the input laser is $P_0$. With 0 degree polarized light (red curve), about 1.5% of the incident energy couples to the grating and into SPPs and starts propagating in the +$x$-direction at around $x = 1500\ nm$. During the propagation, most of that power dissipates, so that only 0.3% of the original incident power reaches the junction at $x = 5900$ nm and couples to the local gap modes. With 90 degree polarized light (blue curve), almost no energy is coupled into SPPs, and therefore almost no energy reaches the junction. For comparison, the calculated the energy absorbed by the nanowire and nanogap under direct excitation of the nanogap with 90 degree polarization, is 2.4% of $P_0$, indicated by the red dashed line. More details about the simulation are discussed in the Supporting Information Sect. S4 and S6. Based on the simulation, therefore, the total coupling efficiency of power to the nanogap just from plasmon dynamics is $\frac{0.3}{2.4} = 12.5\%$ and qualitatively consistent with the 10% experimental result considering the linear relationship between laser power and SERS intensity in our low power region. A number of additional factors can contribute to the slightly smaller experimental result, including inefficiencies in the coupling between propagating SPPs and nanogap LSPRs.

Given the propagation losses, the robustness of the 10% efficiency across many devices implies that the coupling of propagating SPP modes and the nanogap LSPRs is routinely quite efficient, despite microscopic configurational variations from device to device. From the hybridization picture of plasmons[49] this is to be expected. This is in analogy to the idea that a 1D delta function has spectral content from the entire continuum of delocalized harmonic waves in 1D; there should always be overlap and hence coupling between the delocalized SPPs and the highly localized LSPRs at the nanogap.

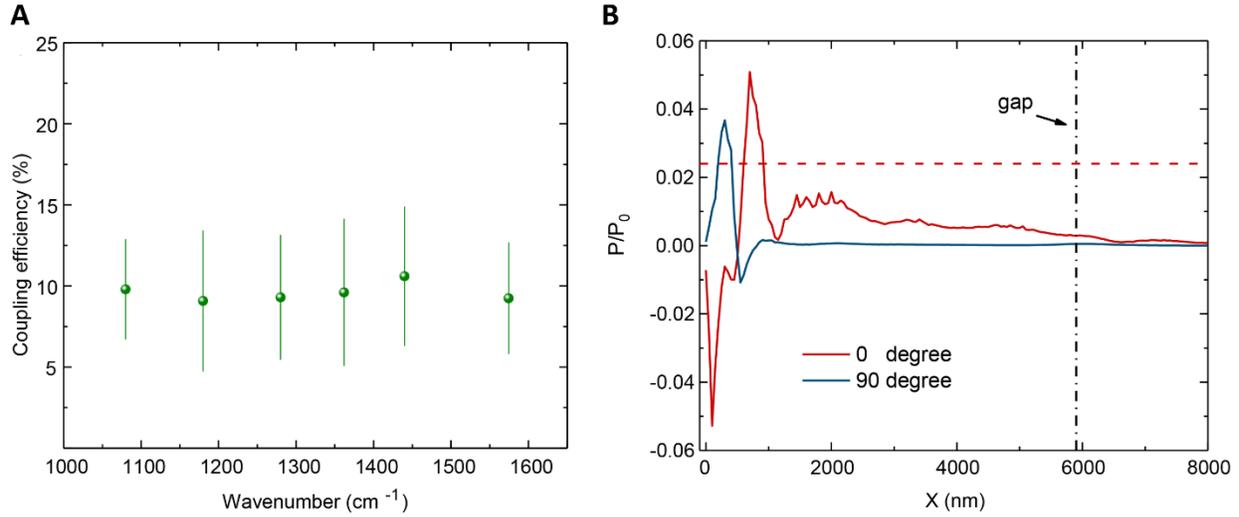

**Figure 3**. (A) Remote SERS coupling efficiency averaged by all devices for each Raman mode. The coupling efficiency is about 10% for all Raman modes. (B) Simulated energy propagation from the grating to the junction the 90 and 0 degree polarization, determined from the local Poynting vector integrated across a cross-section in the *y-z* plane (details in Fig. S9). The *x* origin is at the middle of the grating's two slits. The junction is located at about $x \approx 5900$ nm indicated by the black dashed line. The red dashed line at $\frac{P}{P_0} = 2.4\%$ shows the calculated energy absorption by the junction under direct excitation.

After the map scans, the laser spot can be moved to the pixel where the SERS signal is the largest, and then time dependent spectra are measured by both remote and direct excitation for the same device to check the relative stability of remote SERS, as shown in Figure 4. The input laser intensities are adjusted to make the SERS photon counts comparable. The results shown in Figure 4 are typical and have been seen in multiple devices. For direct excitation, the spectra show temporal intensity fluctuations (blinking) and spectral diffusion in the time range of 200 seconds, which are commonly seen in few or single-molecule SERS.[28,51] Initially, several Raman modes in the range of 1200 to 1400 cm$^{-1}$ are strong and overlap each other. Some modes fade away in 60 seconds and we can only see three modes between 60 to 100 seconds. At 110 seconds, a new mode appears and lasts for about 70 seconds. After 180 seconds, there's no detectable Raman modes, indicating that the nanogap molecular junction has degraded. Previous studies report that the origin of the SERS fluctuation are thermal induced molecular reorientation and sub-nanometer configuration changes of the junction.[28,52,53] In our nanowire device, the thermal heating by direct laser illumination and the localized strong electric field enhancement can contribute to the reconfiguration of the molecules and the Au atoms of the junction, and the fluctuations in spectra are usually accompanied by changes in device resistance.[29,54]

Conversely, the SERS signals are more stable in a much longer time range under remote excitation, even when the photon count rate is approximately the same as the direct excitation case. The strongest Raman modes in 1200 to 1400 cm$^{-1}$ remain stable more than 2500 seconds both for intensity and peak position. There are some fluctuations for other weaker modes, but the magnitude of these fluctuations is smaller than the direct excitation configuration. This is consistent with less thermal heating and more stable junction configurations in the remote excitation configuration, implying that remote excitation methods may be better suited for delicate chemical and biological sensing applications.

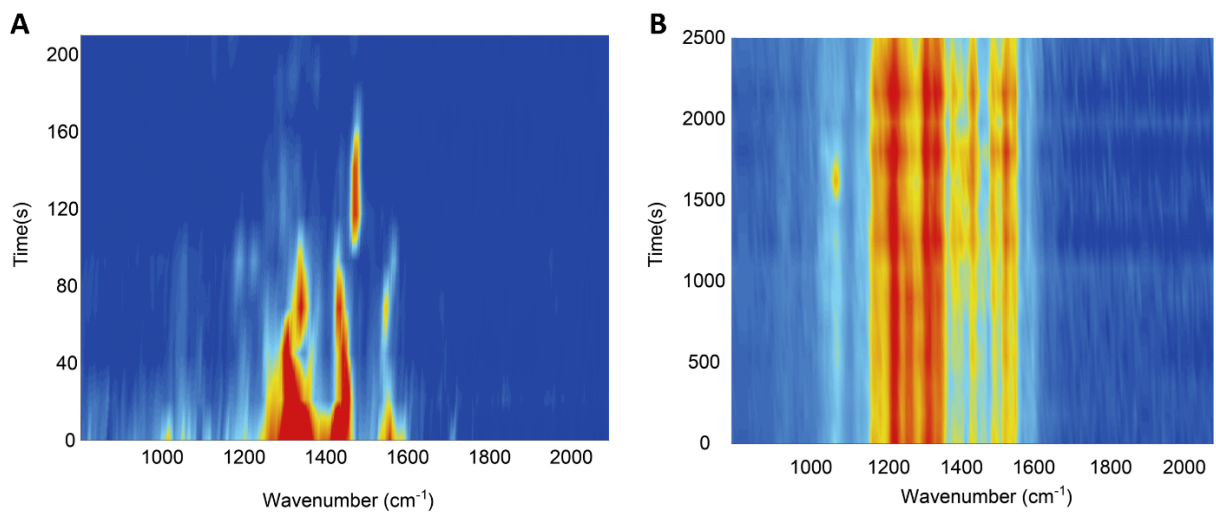

**Figure 4**. Time dependent spectra for the same device by direct and remote excitation. (A) Direct excitation SERS for 200 seconds. (B) Remote excitation SERS for 2500 seconds. The photon count rates are adjusted to be comparable by changing the laser power. Remote excitation SERS is much more stable.

Using nanogap structures that incorporate plasmonic gratings in the electrode design, we have performed a detailed comparison of SERS response under direct excitation (laser incident directly on the nanogap) and remote excitation (laser incident in a grating to launch SPPs toward the nanogap). Across an ensemble of 33 devices, the relative efficiency of remote SERS excitation is about 10%. Through finite element simulations of the electrodynamics of these devices, we find that this is consistent with SPP decay during propagation as the largest source of energy loss. The temporal stability of nanogap SERS response is considerably better under remote excitation than direct excitation, even when the SERS count rate is comparable. This is consistent with the idea[40] that remote excitation significantly reduces local lattice heating at the nanogap site. Remote excitation and its resulting greater junction stability improve the prospects for both sensitive SERS detection of analytes in delicate environments and for combining SERS with challenging electronic spectroscopy techniques such as field-effect gating and inelastic electron tunneling spectroscopy.

## ASSOCIATED CONTENT

The Supporting Information:

Details about device fabrication and measurement setup. Details and discussion about the simulation results. Additional results on the statistical analysis of the ensemble of devices.

## Author information

**Corresponding Author**

**Douglas Natelson** – Department of Physics and Astronomy, Department of Electrical and Computer Engineering, Department of Materials Science and NanoEngineering, Rice University, Houston, Texas 77005, United States; Email: natelson@rice.edu

**Authors**

**Shusen Liao** - Applied Physics Graduate Program, Smalley-Curl Institute, Rice University, Houston, Texas 77005, United States; Department of Physics and Astronomy, Rice University, Houston, Texas 77005, United States

**Yunxuan Zhu** - Department of Physics and Astronomy, Rice University, Houston, Texas 77005, United States

**Qian Ye** - Department of Physics and Astronomy, Rice University, Houston, Texas 77005, United States

**Stephen Sanders** - Department of Electrical and Computer Engineering, Rice University, Houston, Texas 77005, United States

**Jiawei Yang -** Department of Physics and Astronomy, Rice University, Houston, Texas 77005, United States

**Alessandro Alabastri** - Department of Electrical and Computer Engineering, Rice University, Houston, Texas 77005, United States

**Author Contributions**

S.L. and Y.Z. contributed equally to this work.

**Notes**

The authors declare no competing financial interest.

# Acknowledgments

D.N., S.L., and Y.Z. acknowledge Robert A. Welch Foundation grant C-1636 for support of this work. D.N. and J.Y. also acknowledge ONR award N00014-21-1-2062.

# Quantifying Efficiency of Remote Excitation for Surface Enhanced Raman Spectroscopy in Molecular Junctions

Shusen Liao[1,2,†], Yunxuan Zhu[2,†], Qian Ye[2], Stephen Sanders[3], Jiawei Yang[2], Alessandro Alabastri[3], Douglas Natelson[2,3,4*]

[1] Applied Physics Graduate Program, Smalley-Curl Institute, Rice University, Houston, TX 77005 USA

[2] Department of Physics and Astronomy, Rice University, Houston, TX 77005 USA

[3] Department of Electrical and Computer Engineering, Rice University, Houston, TX 77005 USA

[4] Department of Materials Science and NanoEngineering, Rice University, Houston, TX 77005 USA

[†] These authors contributed equally to this work.

# Supporting Information

1. **Device fabrication**
2. **Measurement setup**
3. **Theoretical and simulated analysis on SPPs in vacuum-Au-SiO$_2$ structure**
4. **COMSOL model for direct and remote excitation**
5. **Simulated E field and charge distribution for direct and remote excitation**
6. **Simulated energy flow**
7. **Statistical analysis on coupling efficiency and SERS-OCPV correlation**

---

[*] natelson@rice.edu

## 1. Device fabrication

All the devices are fabricated on Si wafer with 2 μm thick thermal oxide SiO2 layer. First, 50 nm thick large contact Au pads with a 5 nm Ti adhesion layer are prepared by shadow mask E-beam evaporation. The chip is cleaned by acetone, ethanol and isopropyl alcohol and then followed by plasma clean for 5 minutes. After cleaning, two layers of e-beam resist polymethyl methacrylate (PMMA) 495 and 950 are spin coated. PMMA 495 is first spin coated with 3000 rpm for 60 seconds and then PMMA 950 is coated with 4000 rpm for 40 seconds. The nanowire and the gratings with geometry discussed in the main text are written by the Elionix e-beam lithography system. After developing, 30 nm thick Au is evaporated, followed by the lift-off. For molecular self-assembly, the chip is soak in 20 mL ethanol solution with 1 mM BDT for 24 hours inside a nitrogen glovebox. The BDT molecules can be self-assembled on the Au nanowire surface due to Au-S linkage.[1,2] The devices are wire bonded at the Au contact pads and chip carrier after molecule deposition. The nanowire is electromigrated at 30 K to form the junction:[3] an increasing voltage bias is applied to the nanowire and the current is monitored as a feedback simultaneously. The bias increases by step until the current drops, indicating an increase in resistance. Then the bias is set to 0 to start a new cycle. The cycles are repeated until the conductance of the device is smaller than conductance quantum $G_0$. There's chance for the BDT molecules on the Au surface to fall into the nanogap during the electromigration process.

## 2. Measurement setup

All measurements are performed using a home-made Raman system. The device is placed in a Montana Instruments Cryostation and then cooled down to 30 K. We use a 785 nm continuous wave (CW) laser to excite the system. The power of the laser is controlled by a half-wave plate (HWP) and a polarizer. The polarization of the laser is controlled by another HWP. The laser is modulated by an optical chopper, whose frequency is the reference for the lock-in amplifier. The laser is focused at the sample plane by a Zeiss Epiplan-Neofluar 50× objective. The diameter of the laser spot is 1.8 μm measured by knife-edge. We can shine the laser on the junction with normal incidence for direct excitation and shine the laser at the grating with a small angle deviation by tilting mirror M1 after normal incidence. The OCPV is first amplified by the voltage amplifier SRS 560 (not shown in Figure 1B) and measured by the lock-in amplifier SRS DSP 7270. Both remote and direct Raman signals are collected by the same objective. The 785 nm signal reflected by the sample is blocked by the BNF, and the Raman signals are focused at the Synapse CCD spectrometer. We emphasize that the gratings also behave as the SERS substrate, so it is important to block the SERS signals from the grating during SERS measurements by remote excitation. There's an angle difference between the SERS photons coming from the junction and grating when they enter the spectrometer. By setting the slit width of the spectrometer to be 0.05 mm, we can block almost all signals from the grating. This can be further proved by the remote SERS measurement before the electromigration, and we see no Raman mode (Figure 1C). The position sample stage is controlled by a nanopositioner (ANC 300 Piezo Controller) in x, y and z directions. The map scans (Figure 2) are measured with the moving sample stage and fixed laser spot position. A CCD shows the optical image of the device and helps to focus the laser at the junction and gratings.

## 3. Theoretical and simulated analysis on SPPs in vacuum-Au-SiO$_2$ structure.

We employ a simplified 2D model to describe the propagating SPPs. The scheme is shown in Figure S1. The Au layer (region 1) is 30 nm thick and the vacuum (region 3) and SiO$_2$ (region 2) are set to be infinitely thick. Assuming SPPs propagating in x direction, we can get the electric and magnetic field in the three regions:[4]

Region 3:

$$H_y = Ae^{i\beta x}e^{-k_3 z}$$

$$E_x = iA \frac{1}{\omega \varepsilon_0 \varepsilon_3} k_3 e^{i\beta x} e^{-k_3 z}$$

$$E_z = -A \frac{\beta}{\omega \varepsilon_0 \varepsilon_3} e^{i\beta x} e^{-k_3 z}$$

Region 1:

$$H_y = C e^{i\beta x} e^{k_1 z} + D e^{i\beta x} e^{-k_1 z}$$

$$E_x = -iC \frac{1}{\omega \varepsilon_0 \varepsilon_1} k_1 e^{i\beta x} e^{k_1 z} + iD \frac{1}{\omega \varepsilon_0 \varepsilon_1} k_1 e^{i\beta x} e^{-k_1 z}$$

$$E_z = C \frac{\beta}{\omega \varepsilon_0 \varepsilon_1} e^{i\beta x} e^{k_1 z} + D \frac{\beta}{\omega \varepsilon_0 \varepsilon_1} e^{i\beta x} e^{-k_1 z}$$

Region 2:

$$H_y = B e^{i\beta x} e^{k_2 z}$$

$$E_x = -iB \frac{1}{\omega \varepsilon_0 \varepsilon_2} k_2 e^{i\beta x} e^{k_2 z}$$

$$E_z = -B \frac{\beta}{\omega \varepsilon_0 \varepsilon_2} e^{i\beta x} e^{k_2 z}$$

$k_1, k_2, k_3$ are the wave vectors in region 1, 2, 3 in z direction respectively. $\beta$ is the wave vector in SPPs propagation direction. $\varepsilon_1, \varepsilon_2, \varepsilon_3$ are the dielectric function in region 1, 2, 3 respectively. $\omega$ is the frequency of the input light.

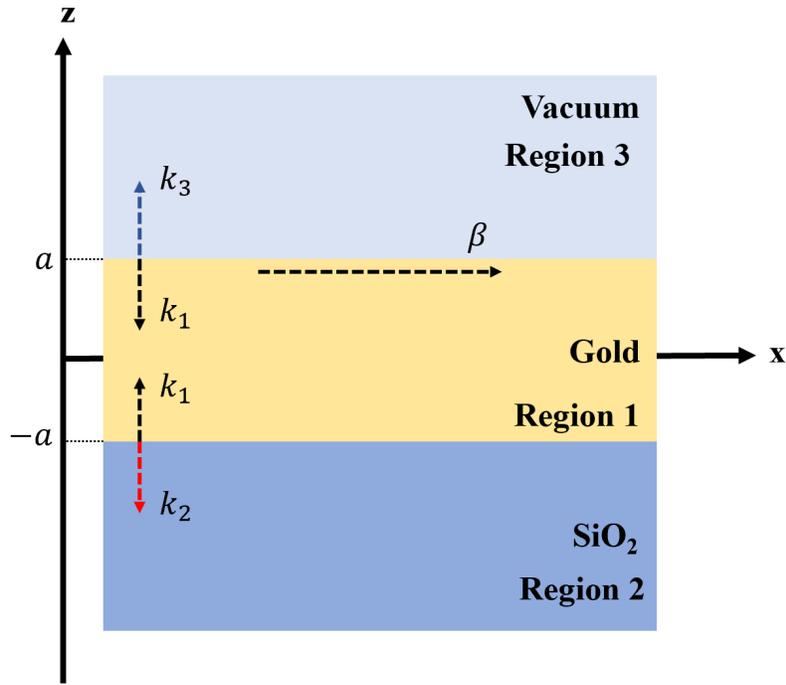

**Figure S1**. Scheme of the SPPs mode in Vacuum-Au-SiO$_2$ structure, considering TM modes and SPP propagation in the x direction.

Consider the continuity of $H_y$ and $E_x$, we have:

$$Ae^{-k_3 a} = Ce^{k_1 a} + De^{-k_1 a}$$

$$\frac{A}{\varepsilon_3}k_3 e^{-k_3 a} = -\frac{C}{\varepsilon_1}k_1 e^{k_1 a} + \frac{D}{\varepsilon_1}k_1 e^{-k_1 a}$$

$$Be^{-k_2 a} = Ce^{-k_1 a} + De^{k_1 a}$$

$$-\frac{B}{\varepsilon_2}k_2 e^{-k_2 a} = -\frac{C}{\varepsilon_1}k_1 e^{-k_1 a} + \frac{D}{\varepsilon_1}k_1 e^{k_1 a}$$

$2a = 30\ nm$ is the thickness of Au. Then we get the dispersion relation:

$$e^{-4k_1 a} = \frac{k_1/\varepsilon_1 + k_2/\varepsilon_2}{k_1/\varepsilon_1 - k_2/\varepsilon_2} \cdot \frac{k_1/\varepsilon_1 + k_3/\varepsilon_3}{k_1/\varepsilon_1 - k_3/\varepsilon_3} \quad (1)$$

The wave equation for TM modes is:

$$\frac{\partial^2 H_y}{\partial z^2} + (k_0^2 \varepsilon - \beta^2) H_y = 0$$

We have:

$$k_i^2 = \beta^2 - k_0^2 \varepsilon_i \quad i = 1, 2, 3 \quad (2)$$

$k_0$ is the wave vector of the input laser. The values of dielectric functions are determined by the material:[5] $\varepsilon_1 = -20.669 + 1.4246i$, $\varepsilon_2 = 2.113$, $\varepsilon_3 = 1$. By solving equations (1) and (2), we get every wave vector in this system:

$$k_1 = 3.858 \times 10^7 + 1.139 \times 10^6 i$$

$$k_2 = 5.25 \times 10^6 + 3.256 \times 10^5 i$$

$$k_3 = 9.94 \times 10^6 + 1.720 \times 10^5 i$$

$$\beta = 1.276 \times 10^7 + 1.339 \times 10^5 i$$

The wavelength of SPPs is calculated to be $Re\left(\frac{2\pi}{\beta}\right) = 492$ nm. A simulation of SPPs generation is performed for comparison. The simulation is performed in COMSOL with a 2D model containing the gratings (Figure S2A). The simulated wavelength of the SPPs is 489 nm (Figure S2B), which agrees well with the theoretical result. The diffraction pattern through the grating slits is clearly seen in Figure S3B.

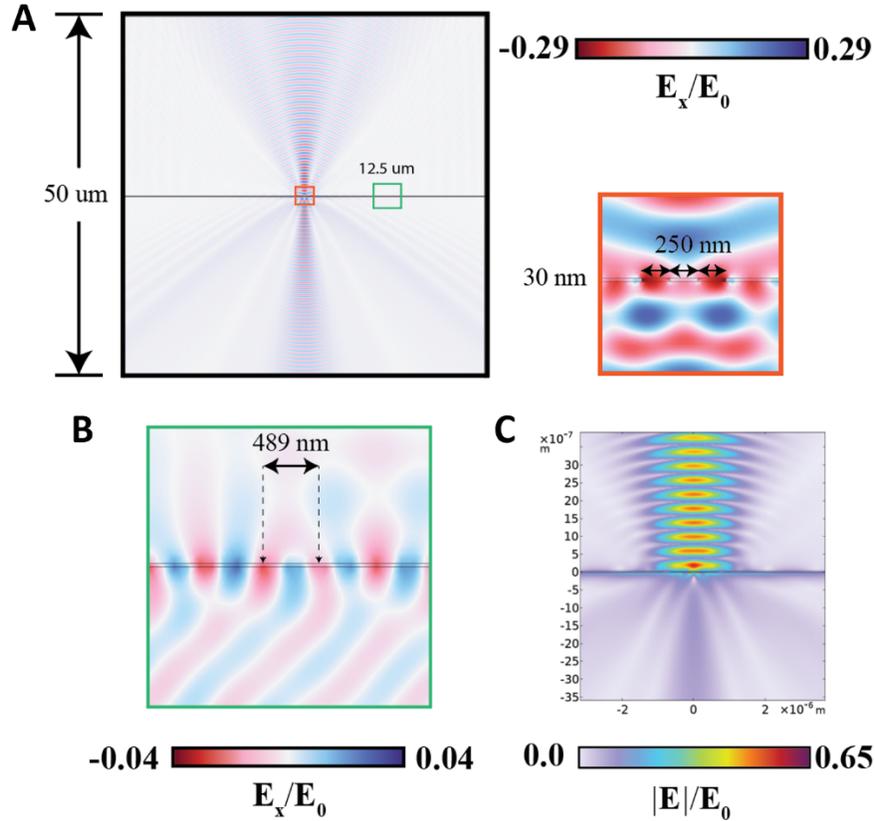

**Figure S2**. Simulation of SPP generation. (A) Scheme of the 2D model with Gaussian laser beam input, a cross-section of the *x-z* plane, with light incident from +*z* and SPP propagation in the *x* direction. The right panel (the enlarged image of the orange box in the left panel) shows the grating geometry and the surrounding electric field distribution. The Au layer is 30 nm thick. (B) Electric field distribution in the area indicated by the green box in (A). The wavelength of the SPPs is 489 nm. (C) Electric field amplitude distribution. A diffraction pattern is formed in the substrate.

To better understand of the SPPs in our system, we simulate the SPPs at the Au-vacuum and Au-SiO$_2$ interfaces. The results are shown in Figure S3. The decay length of the electric field in *z* direction is about 30 nm for both cases, which means the SPPs in the vacuum-Au-SiO$_2$ structure are the summation of SPPs in Au-vacuum and Au-SiO$_2$ interface.

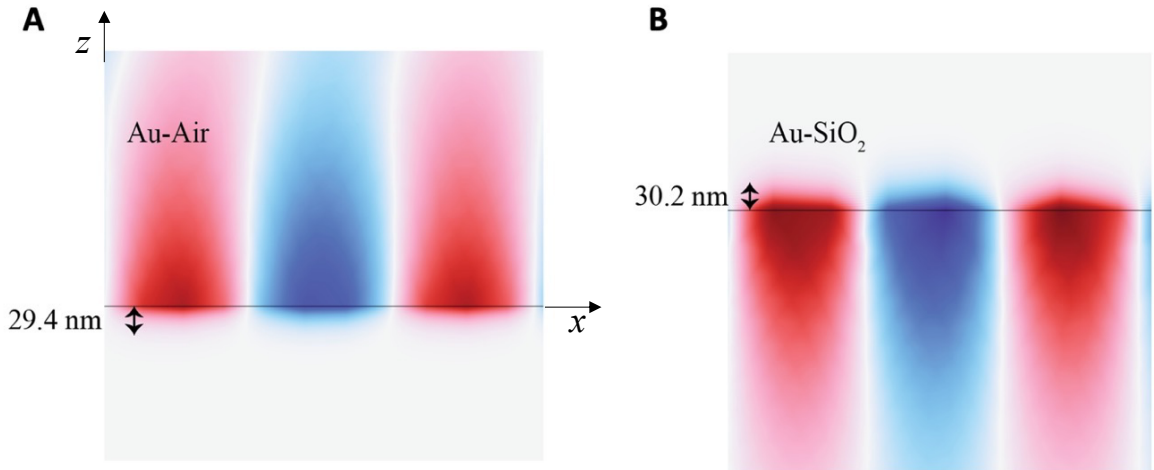

**Figure S3**. (A) Simulated SPPs at Au-vacuum interface, the decay length in Au is 29.4 nm, view in the *x-z* plane. (B) Simulated SPPs at Au-SiO$_2$ interface, the decay length in Au is 30.2 nm. The vacuum and SiO$_2$ are set to be infinitely thick.

## 4. COMSOL model for direct and remote excitation.

A finite element method (FEM) simulation is perfomed using COMSOL Multiphysics 6.1. The geometry of the model (Figure S4) is defined as in the description in the main text. To simplify the model, we only consider the grating in one of the large electrodes. The model is surrounded by perfect match layers (PML). The simulation is performed in the wave optics module electromagnetic wave frequency domain (ewfd) section with two steps: the first step defines the background Gaussian beam input laser, and the second step calculates the scattering electric field.

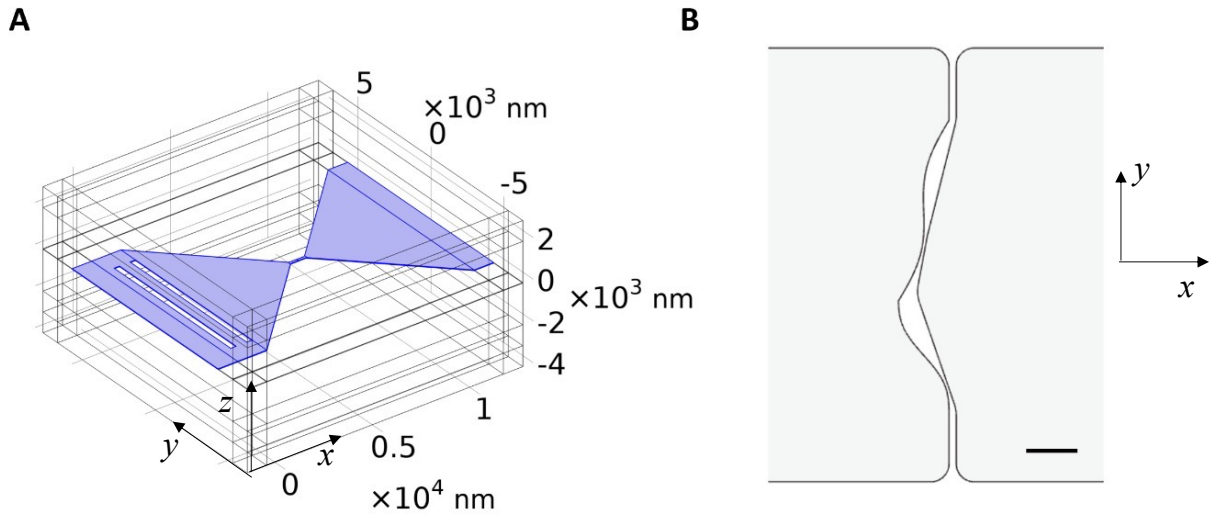

**Figure S4**. (A) Geometry of the whole model. The Au layer is colored in blue. (B) The geometry of the junction gap. Scale bar is 15 nm.

## 5. Simulated E field and charge distribution for direct and remote excitation.

The electric charge and field distribution around the gap can be calculated as shown in Figure S5 and S6 for remote and direct excitation, respectively. In the case of *remote* excitation, both the electric field and charge at the junction for 0 degree (*x*) polarized light are larger than 90 degree (*y*) polarized light, which qualitatively agrees well with the experimental results in Figure 2. This also makes qualitative sense, as polarization across the grating is expected to excite propagating SPPs. We still can see the simulated field enhancement factor exceeding 20 for 90 degree polarized light because of the contribution of diffracted light propagating in the $SiO_2$ substrate to the gap. However, when it comes to direct excitation, we can see the electric field is greatly enhanced for both polarizations. The enhancement factors are over 100, much larger than remote excitation. This agrees well with the experimental results in Figure 2.

For direct excitation, however, both the electric field and charge for 0 degree (*x*) polarized light are larger than for 90 degree (*y*) polarized light, which is inconsistent with the experimental results. By reducing the thickness of Au layer in the simulation to 20 nm, we find results consistent with the experiments, with larger enhancement factor and electric charge under 90 degree polarized light excitation (Figure S7). These simulations show that the polarization dependence of the enhanced field is quite sensitive to the assumed thickness of the metal. There are several potential contributors to this issue. First, the effective dielectric function of the real Au layer may deviate slightly from the theory value we use in the simulation due to adsorbates, Au surface roughness and grain boundaries, and other defects from device fabrication. Second, the thickness of the Au layer may deviate from the designed value of 30 nm by a few nm during e-beam evaporation, and the local thickness of the metal near the junction can change during the electromigration process. Third, the polarization dependence behavior is also related to the detailed geometry of the gap, which can evolve through annealing between when the measurements are performed and when electron microscopy is used to find a geometry to use in the simulation. The relationship between the Au thickness in simulation and field enhancement factor for both polarizations are shown in Figure S8.

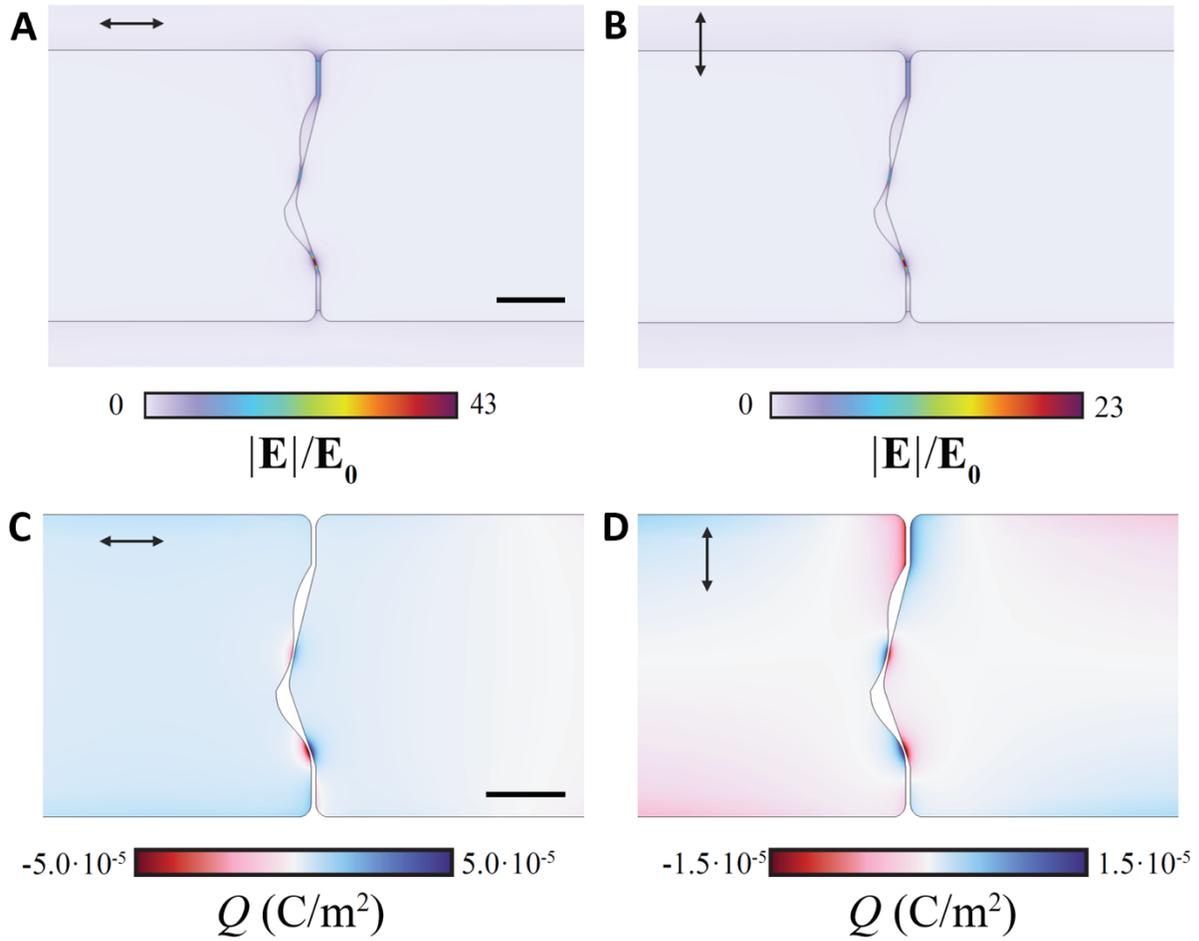

**Figure S5**. Electric field enhancement and charge distribution at the nanogap under *remote* excitation (laser incident on grating) with 30 nm thick Au layer. (A), (B) Field enhancement with 0 degree (*x*-aligned) and 90 degree (*y*-aligned) laser polarizations, respectively. (C) (D) Electric charge distributions 0 degree (*x*-aligned) and 90 degree (*y*-aligned) laser polarizations, respectively. Polarization directs are indicated by the black arrows. The scales bars are 30 nm.

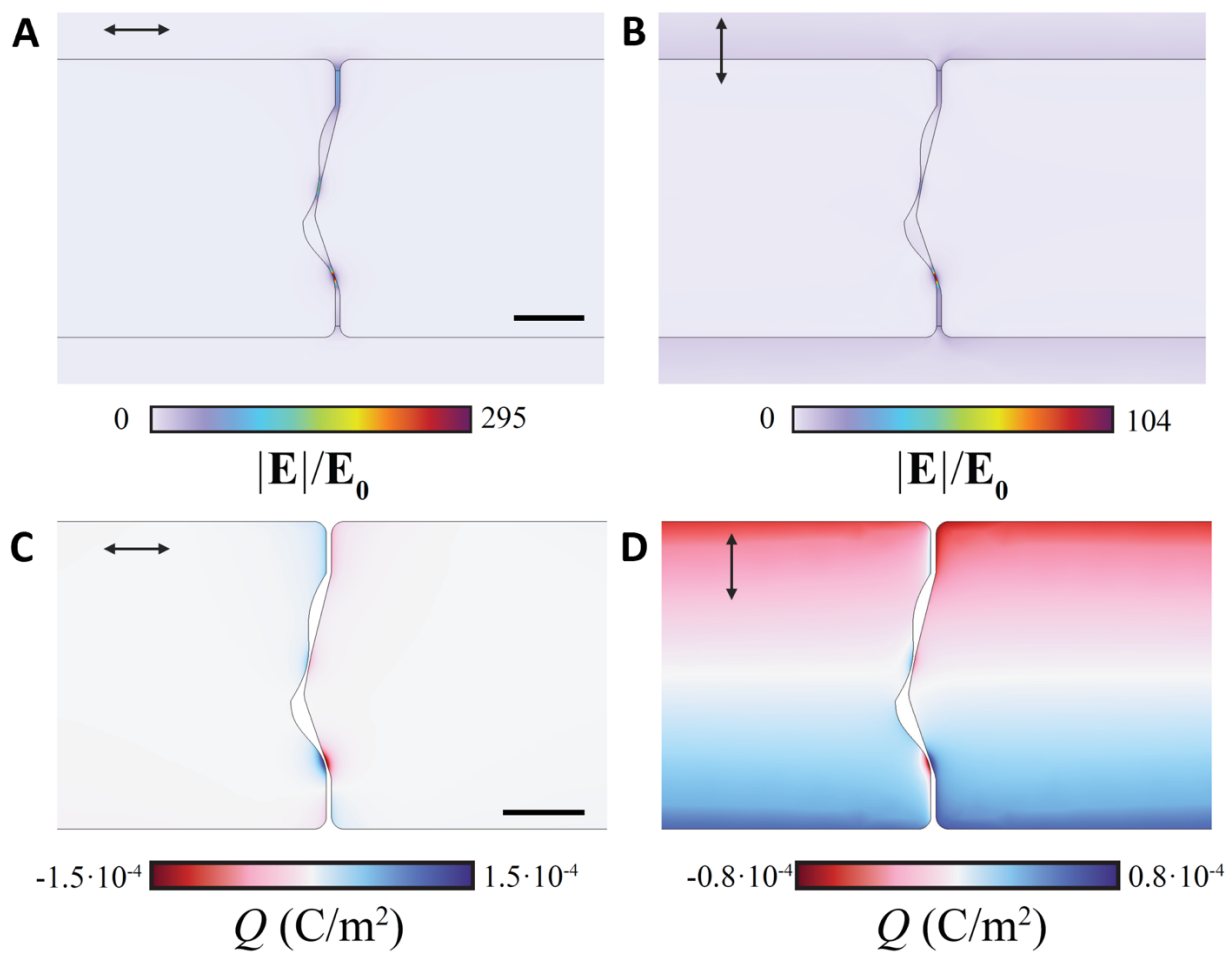

**Figure S6**. Electric field enhancement and charge distribution at the nanogap under *direct* excitation with 30 nm thick Au layer. (A) (B) Field enhancement with 0 degree (*x*-aligned) and 90 degree (*y*-aligned) laser polarizations, respectively. (C) (D) Electric charge distribution with 0 degree (*x*-aligned) and 90 degree (*y*-aligned) laser polarizations, respectively. Polarization directs are indicated by the black arrows. The scales bars are 30 nm.

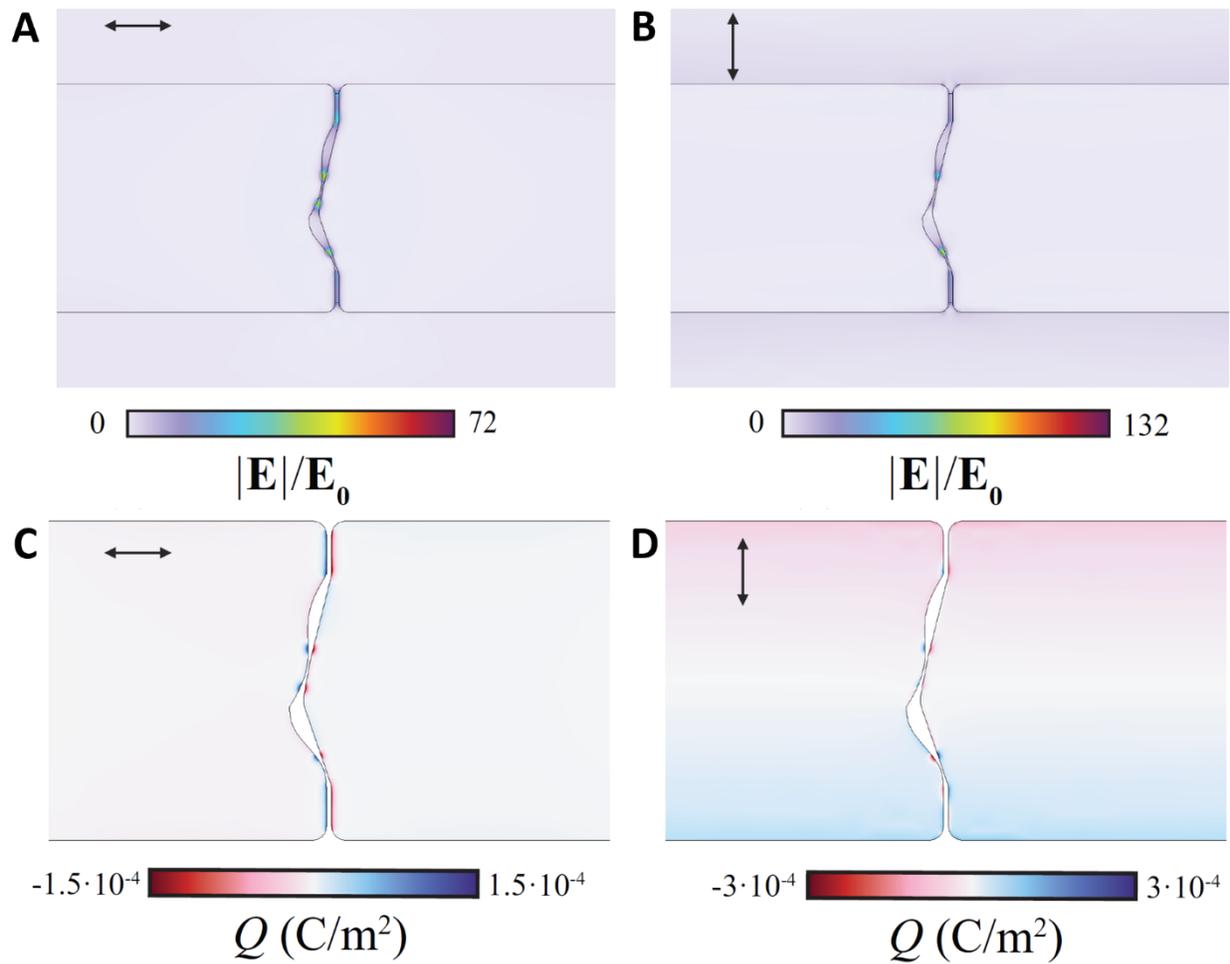

**Figure S7**. Electric field enhancement and charge distribution at the nanogap under *direct* excitation with 20 nm thick Au layer. (A) (B) Field enhancement with 0 degree (*x*-aligned) and 90 degree (*y*-aligned) laser polarizations, respectively. (C) (D) Electric charge distribution with 0 degree (*x*-aligned) and 90 degree (*y*-aligned) laser polarizations, respectively. Polarization directs are indicated by the black arrows. The scales bars are 30 nm.

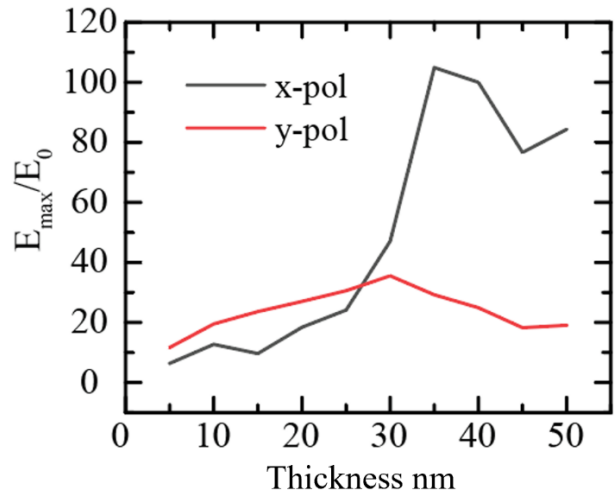

**Figure S8**. The maximum of electric field enhancement with *x* (0 degree) and *y* (90 degree) polarized light in the gap under direct illumination of the nanogap as a function of Au thickness.

## 6. Simulated energy flow.

The energy flow from the grating to the junction (Figure 3B) is calculated by integrating the Poynting vector along *x* (propagation) direction over an area in the *y-z* plane, as shown in Figure S9. The rectangular area contains the whole Au wire in the *y* direction. In the *z* direction, the height is set to be 130 nm to include the 30 nm Au and 50 nm in both vacuum and $SiO_2$. The 50 nm is chosen to include most of the SPPs as well as to exclude light reflected by the gold layer.

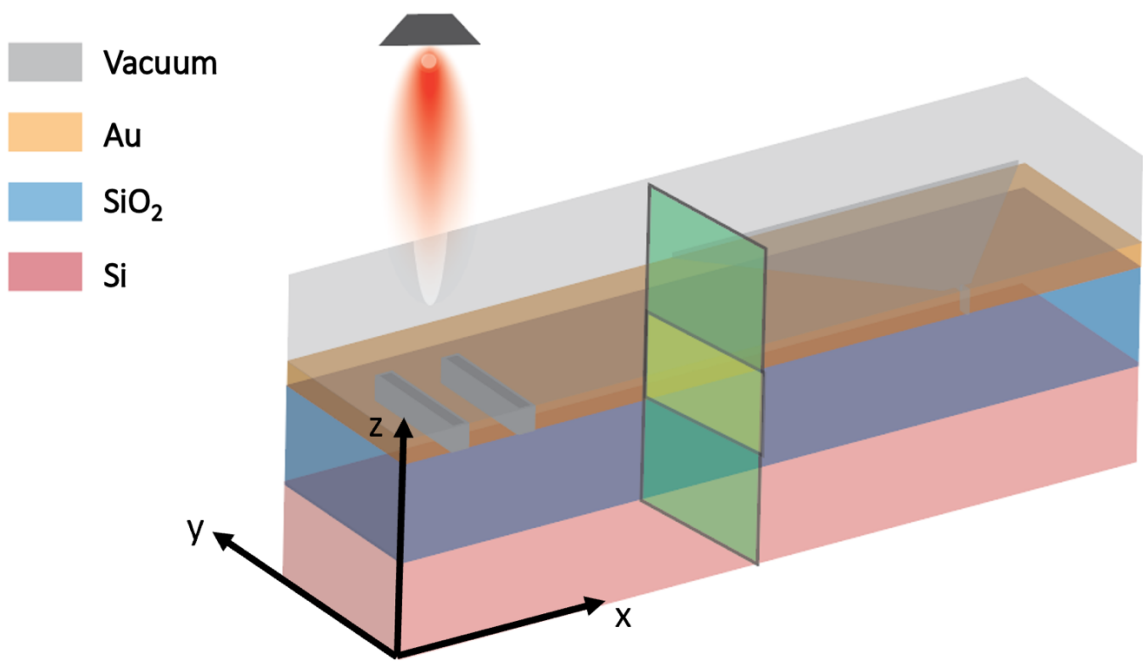

**Figure S9**. Scheme of energy flow calculation. The *x* component of the Poynting vector is integrated over the area indicated by the yellow rectangular in *y-z* plane.

## 7. Statistical analysis on coupling efficiency and SERS-OCPV correlation

As described in the main manuscript, we can define a SERS coupling efficiency by comparing the SERS count rates at a given incident power for remote and direct excitation of the nanogap. Equivalently, we can measure what power is required in the remote excitation configuration to produce the same number of SERS counts as in the direct excitation geometry. For a single device, we can calculate the coupling efficiency by averaging over all Raman modes. The results for the ensemble of devices are shown in Figure S10. For most of devices, the coupling efficiency is around 10%. However, we can still see some devices that show low coupling efficiency, less than 5%. This may be caused by defects in the Au layer that affect the propagation of SPPs, or by particular nanogap geometries that have lower coupling between the propagating SPPs and the nanogap LSPRs.

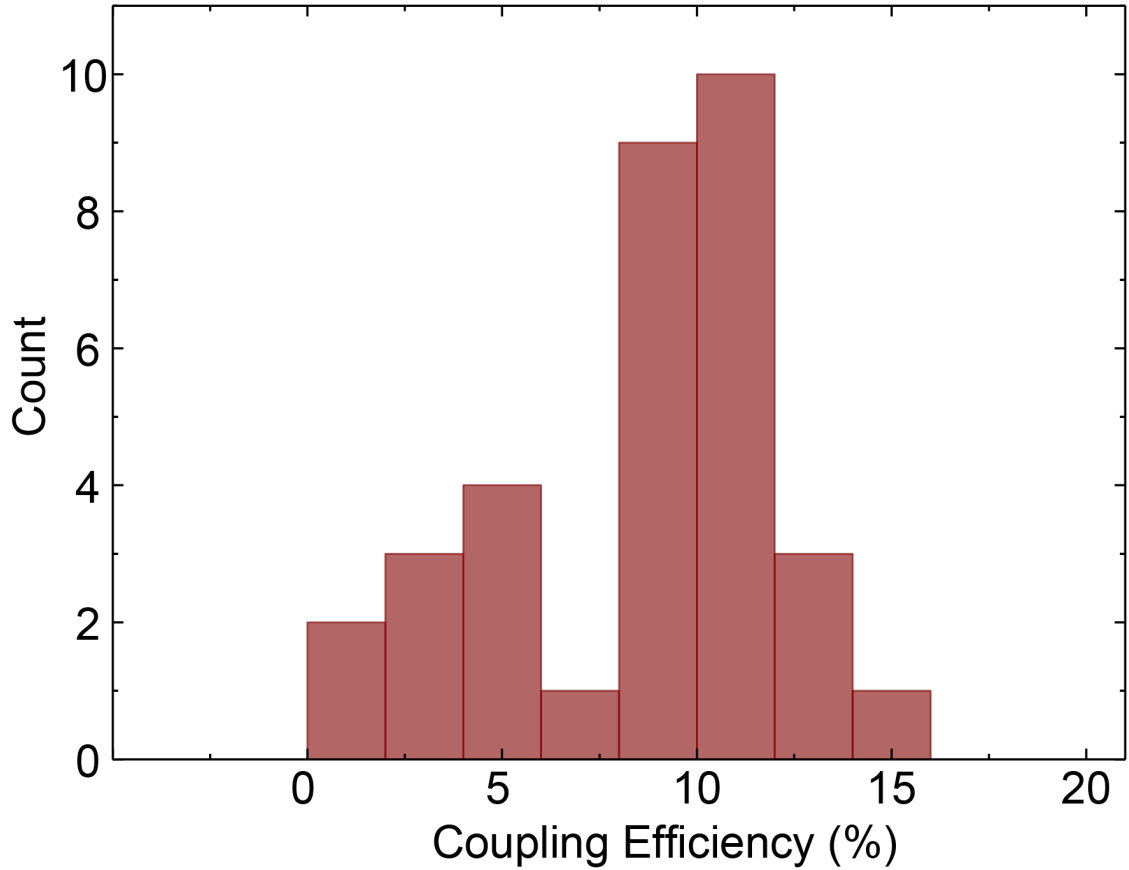

**Figure S10**. Histogram of coupling efficiency averaged by all Raman modes for each device.

As mentioned in the manuscript, the correlation between OCPV and SERS can be complicated. The correlation is shown in Figure S11 for all the devices under both direct and remote excitation. The SERS signal is the summation of photon of all Raman mode in the range of 1000-1600 cm$^{-1}$. The hot carrier current per incident optical power for a single device is defined as:[6]

$$I_{hot} = \frac{OCPV}{R \cdot P_{laser}}$$

$R$ is the resistance of the device; $P_{laser}$ is the laser power. The hot carrier current involves the rate of plasmon-decay-based generation of carriers as well as asymmetries in the junction that lead to a dominant direction for carrier tunneling[6]. This normalization by laser power is reasonable as the OCPV is observed in the experiments to be linear in the incident power. Using the current rather than the OCPV normalizes out differences in device resistances.

Strong excitation of nanogap LSPRs is essential to generate a substantial OCPV (hot carrier current), just as such LSPR excitation is needed for SERS response. As a result, for both direct and remote excitation, we can see a rough positive correlation between hot electron current and SERS. As described in the manuscript, however, the magnitude of the hot carrier current also depends on the asymmetry of the nanogap and the particular LSPR modes. For example, we should have very small hot carrier current for devices with highly symmetric gap geometry because the tunneling current from the two sides of the nanogap would roughly cancel, even if the plasmon excitation is strong. Such a device may still produce a strong SERS signal due to the large electric field enhancement by plasmon resonance. This introduces fluctuations into the rough positive correlation, as seen in Fig. S11. Conversely, we can hardly have devices with large hot carrier current but weak SERS signal, because devices with large hot carrier current must have strong plasmon response which contributes to strong SERS signal.

We note that the OCPV in the nanogaps is not a photothermoelectric voltage, as discussed elsewhere [6]. There is some amount of local heating associated with the hot carriers responsible for the OCPV, since the excess energy of those carriers eventually ends up in the phonons of the lattice and spread throughout the Fermi sea. That being said, light emission experiments[7] have established that hot carriers produced at the nanogap due to plasmon excitation and decay (electrically[7] or optically[8]) are, in the steady state, a small fraction of the total Fermi sea and the lattice remains much colder than those carriers. Direct measurements of local electronic and lattice temperature at the nanogap remain challenging.

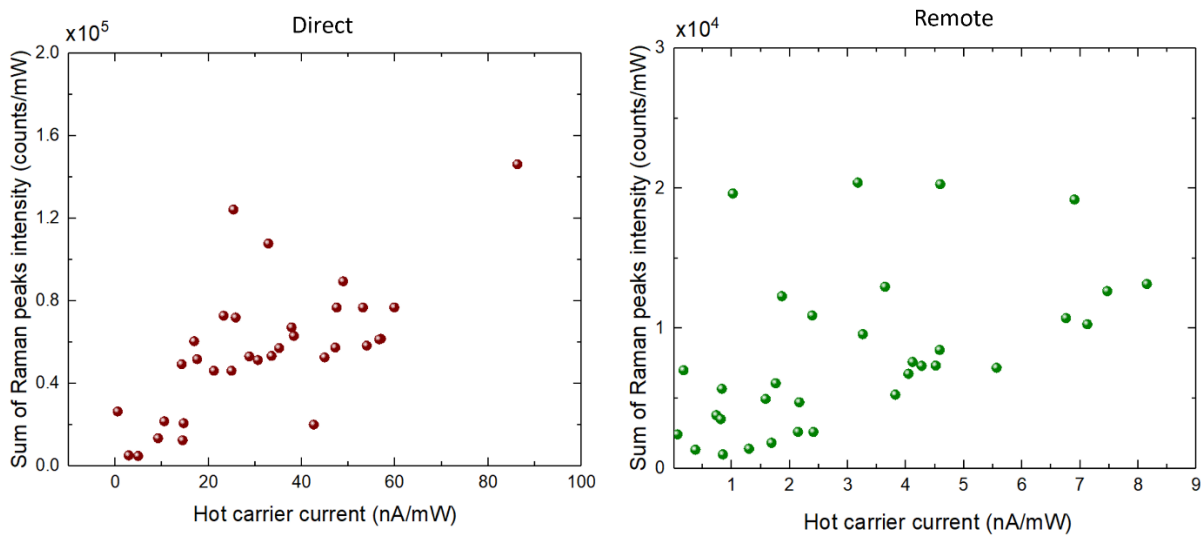

**Figure S11**. Correlation between OCPV (hot carrier current) and SERS by direct and remote excitation.